\documentclass{aastex63}
\usepackage{bm,amsmath}
\usepackage{dcolumn}% Align table columns on decimal point
\usepackage{bm}% bold math
\usepackage{epsf}
\usepackage{verbatim}
\usepackage{amsmath}
\usepackage{amssymb}
\usepackage{color}
\usepackage{threeparttable}
\usepackage{multirow}
\usepackage{verbatim} % needed for comments
\usepackage[normalem]{ulem} % needed for strikeout font
\usepackage{mathtools}
\usepackage{amsfonts}
\usepackage{relsize}

\received{\today}
\revised{}
\accepted{}
\shorttitle{X-raying the Birth of BNSs and NS-BHs}
\shortauthors{Kashiyama et al.}
\graphicspath{{./}{figures/}}
\begin{document}

\title{X-raying the Birth of Binary Neutron Stars and Neutron Star-Black Hole Binaries}

\correspondingauthor{Kazumi Kashiyama}
\email{kashiyama@astr.tohoku.ac.jp}

\author[0000-0003-4299-8799]{Kazumi Kashiyama}
\affiliation{Astronomical Institute, Graduate School of Science, Tohoku University, Aoba, Sendai 980-8578, Japan}
\affiliation{Kavli Institute for the Physics and Mathematics of the Universe (Kavli IPMU,WPI), The University of Tokyo, Chiba 277-8582, Japan}
\affiliation{Department of Physics, Graduate School of Science, University of Tokyo, Bunkyo-ku, Tokyo 113-0033, Japan}
\affiliation{Research Center for the Early Universe, Graduate School of Science, University of Tokyo, Bunkyo-ku, Tokyo 113-0033, Japan}

\author[0000-0003-4876-5996]{Ryo Sawada}
\affiliation{Department of Earth Science and Astronomy, Graduate School of Arts and Sciences, The University of Tokyo, Tokyo 153-8902, Japan}

\author[0000-0002-7443-2215]{Yudai Suwa}
\affiliation{Department of Earth Science and Astronomy, Graduate School of Arts and Sciences, The University of Tokyo, Tokyo 153-8902, Japan}
\affiliation{Center for Gravitational Physics and Quantum Information, Yukawa Institute for Theoretical Physics, Kyoto University, Kyoto 606-8502, Japan}



\begin{abstract}

We consider fallback accretion after an ultra-stripped supernova (USSN) that accompanies formation of a binary neutron star (BNS) or a neutron star-black hole binary (NS-BH). The fallback matter initially accretes directly to the nascent NS, while it starts to accrete to the circumbinary disk, typically $0.1\mbox{-}1\, \mathrm{day}$ after the onset of the USSN explosion. The circumbinary disk mass further accretes, forming mini disks around each compact object, with a super-Eddington rate up to a few years. We show that such a system constitutes a binary ultraluminous X-ray source (ULX), and a fraction of the X rays can emerge through the USSN ejecta. We encourage follow-up observations of USSNe within $\lesssim 100\,\rm Mpc$ and $\sim 100\mbox{-}1,000\,\mathrm{day}$ after the explosion using {\it Chandra}, {\it XMM Newton} and {\it NuSTAR}, which could detect the X-ray counterpart with time variations representing the properties of the nascent compact binary, e.g., the orbital motion of the binary, the spin of the NS, and/or the quasiperiodic oscillation of the mini disks. 

\end{abstract}



\section{Introduction}\label{sec:intro}
The LIGO-Virgo detector network has detected gravitational waves (GWs) from $O(100)$ of coalescing compact binaries~\citep{2021arXiv211103606T}. Most of them are confirmed to be binary black holes (BBHs) except for two binary neutron stars~\citep[BNSs;][]{2017PhRvL.119p1101A,2020ApJ...892L...3A} and two neutron star-black hole binaries~\citep[NS-BHs;][]{2021ApJ...915L...5A}. %The upcoming observation runs could detect $\sim 10\mbox{-}100$ events per year. 
Their stellar progenitors are of great astrophysical interest. 

In the isolated binary evolution scenario, the births of the BNSs and NS-BHs are likely accompanied by ultra-stripped supernovae~~\citep[USSNe;][]{2013ApJ...778L..23T,2015MNRAS.451.2123T}; in order for the compact binary (CB) to lose its orbital energy via GW emission and merge within a cosmological time, the orbital separation at the birth needs to be comparable to the size of a massive stellar core, which means that the progenitor of the second-born NS inevitably experiences a significant envelope stripping via the binary interaction. %This scenario has been supported by binary stellar evolution calculations. 

Properties of such USSNe have been inferred both by theory and observation. Neutrino-radiation hydrodynamic simulations show that a collapse of a carbon oxygen (CO) core is responsible for a successful explosion with an explosion energy of $E_{\rm SN} \sim 10^{50}\,\mathrm{erg}$ and an ejecta mass of $M_\mathrm{ej} \sim 0.1\,M_\odot$~\citep[e.g.,][]{2015MNRAS.454.3073S,2018MNRAS.479.3675M}. See also \cite{2022arXiv220308292M} for the electron-capture SN explosion. Either way, the weak explosion induces a weak natal kick, preventing the disruption of the binary. %and is consistent with the orbital parameters of a population of the Galactic BNSs~\citep[e.g.,][]{2005PhRvL..94e1102P,2007AIPC..924..598V,2016MNRAS.456.4089B}. 
Besides, high-cadence transient surveys are identifying more and more USSN candidates, i.e., those with faster light curves than ordinary core-collapse SNe and spectroscopic signatures of ultra-stripped progenitor~\citep[e.g.,][]{2018Sci...362..201D,2020ApJ...900...46Y}. 

With the increasing USSN samples at hand, the question is {\it ``Which USSNe accompany what type of BNS/NS-BH formation, in particular, those coalesce within a cosmological timescale?"}. In fact, the estimated USSN rate, ${\cal R}_{\rm USSN} \gtrsim {\rm a\,few\,}\times 1,000\,\mathrm{yr^{-1}\,Gpc^{-3}}$~\citep[e.g.,][]{2019ApJ...882...93H}, is an order of magnitude higher than the observed merger rates of BNS and NS-BH, ${\cal R}_{\rm BNS} = 10\mbox{-}1,700\,\mathrm{yr^{-1}\,Gpc^{-3}}$ and ${\cal R}_{\rm NS-BH} = 7.4\mbox{-}320\,\mathrm{yr^{-1}\,Gpc^{-3}}$~\citep{2021arXiv211103634T}. On the other hand, {\it ``What is the energy source of USSNe?"} is another important question regarding the SN mechanism; although the theoretically calculated explosion energy and ejecta mass are consistent with those inferred from observations, the $^{56}$Ni masses given by the same theoretical calculations are at most $\sim 0.01 M_\odot$ and insufficient to explain some USSNe, for example, iPTF14gqr~\citep{2022ApJ...927..223S}. This may indicate that additional energy injection from the nascent NS and/or CB is necessary~\citep{2022ApJ...927..223S}.\footnote{In recent years, it has been noted that the $^{56}$Ni production problem also exists for canonical supernovae \citep{2019MNRAS.483.3607S,2019ApJ...886...47S,2021ApJ...908....6S}.}

\begin{figure}
\centering
  \includegraphics[width=1.\textwidth]{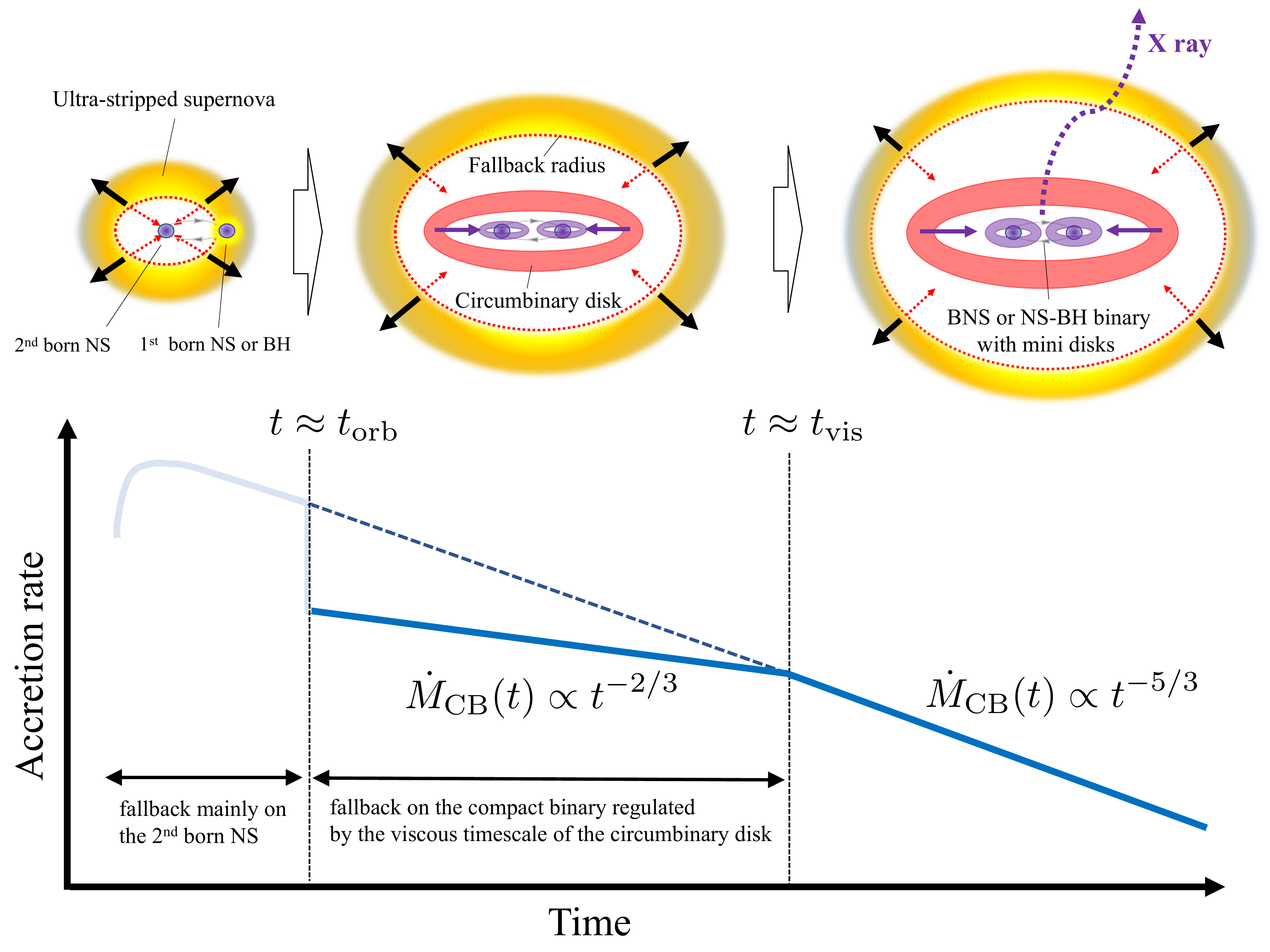} 
\caption{
Schematic picture of ultra-stripped supernova fallback onto a nascent binary neutron star (BNS) or neutron star-black hole binary (NS-BH). 
} 
\label{fig:schem}
\end{figure}

In order to answer the above questions, one needs to have direct evidence of the formation of CB in a USSN and probe their properties, e.g., orbital separation and eccentricity of the binary, magnetic field strength and spin period of the NSs. To this end, we here consider fallback accretion occurring after the USSN and propose to search for the X-ray counterpart (See Fig. \ref{fig:schem}). An orbital timescale after the explosion, the fallback matter should start to accrete to the circumbinary disk and mini disks are also formed around each compact object. We show that the accretion rate well exceeds the Eddington limit for a few years, and the nascent CB can be a binary ultraluminous X-ray source (ULX). The physical properties of the nascent CB are imprinted in the X-ray emission, in particular in its time variations.  

This paper is constructed as follows. In Sec. \ref{sec:ussn}, we review the progenitor system of BNSs and NS-BHs that merge within a cosmological timescale and the USSN explosion associated with the second-born NS formation. In Sec. \ref{sec:fallback}, we theoretically model fallback accretion onto nascent BNSs and NS-BHs. We calculate the resultant X-ray emission in Sec. \ref{sec:xray} and the detectability in Sec. \ref{sec:xray_esc}. Sec. \ref{sec:summary} is for summary and discussion. We use the notation $Q = 10^x\,Q_x$ in CGS units, except for some mass parameters in units of $M = 10^y\,M_y\,M_\odot$.

\section{Ultra-stripped supernova accompanying compact binary formation}\label{sec:ussn}
Let us first review the basic properties of a USSN accompanying formation of a CB that coalesces within a cosmological time. 

\subsection{Progenitor binary system}
The GW insprial time of the CB is calculated as
\begin{equation}
t_{\rm GW} \sim 0.16\,{\rm Gyr}\,a_{\rm 11}^4 \left(\frac{\mu}{0.7\,M_\odot}\right)^{-1}\left(\frac{m}{2.8\,M_\odot}\right)^{-2}(1-e^2)^{7/2},
\end{equation}
where $m = m_1 + m_2$ is the total mass, $\mu = m_1m_2/(m_1+m_2)$ is the reduced mass, $a$ is the semi-major axis, and $e$ is the eccentricity of the binary.
The corresponding orbital period is 
\begin{equation}
t_{\rm orb} \sim 0.1\,{\rm day}\,a_{11}^{3/2} \left(\frac{m}{2.8\,M_\odot}\right)^{-1/2}.
%= 2\pi\left(\frac{a^3}{Gm}\right)^{1/2} 
\end{equation}
Hence, to merge within a cosmological time, say $t_{\rm GW} \lesssim 10\,\mathrm{Gyr}$, the CB orbit at its birth needs to satisfy the following criteria, 
\begin{equation}\label{eq:a_crit}
    a \lesssim 3\,\times 10^{11}\,\mathrm{cm}\,\left(\frac{t_{\rm GW}}{10\,\mathrm{Gyr}}\right)^{1/4}\left(\frac{\mu}{0.7\,M_\odot}\right)^{1/4}\left(\frac{m}{2.8\,M_\odot}\right)^{1/2}(1-e^2)^{-7/8},
\end{equation}
and 
\begin{equation}
    t_\mathrm{orb} \lesssim 0.5\,\mathrm{day}\,\left(\frac{t_{\rm GW}}{10\,\mathrm{Gyr}}\right)^{3/8} \left(\frac{\mu}{0.7\,M_\odot}\right)^{3/8}\left(\frac{m}{2.8\,M_\odot}\right)^{1/4}(1-e^2)^{-21/16}.
\end{equation}

In the isolated binary evolution scenario, such a CB can be formed from an OB star binary with an initial orbital separation of $a \lesssim 1\,\mathrm{AU}$~\citep[see e.g.,][for a review]{2014LRR....17....3P}: In the post-main-sequence phase of the primary, the first common envelope may develop. After spiraling in to some extent, the primary stellar core explodes and forms an NS or a BH. If the natal kick of the first-born compact object is sufficiently small not to significantly enlarge the binary separation, the second common envelope phase may occur. The system evolves into (near) contact binary of the NS/BH and a helium star. In this case, the Case BB Roche robe overflow occurs; the envelope of the helium star is further stripped, and its mass becomes close to the Chandrasekhar mass. The fate of such an ultra-stripped star may be an electron-capture SN or a core-collapse SN driven by the neutrino mechanism~\citep{2015MNRAS.451.2123T}. In this paper, we focus on the latter case (see, e.g., \cite{2022arXiv220308292M} for the former case). 

\subsection{Ultra-stripped supernovae}
Hydrodynamic simulations of core collapse of ultra-stripped progenitors have been performed~\citep{2015MNRAS.454.3073S,2017MNRAS.471.4275Y,2017MNRAS.466.2085M,2018MNRAS.479.3675M,2018MNRAS.481.3305S,2022ApJ...927..223S}. It typically leads to a successful explosion by the neutrino mechanism with an explosion energy of $E_{\rm SN} \sim 10^{50}\,\rm erg$, an ejecta mass of $M_{\rm ej} \sim 0.1\,M_\odot$, and an ejected $^{56}$Ni mass of $M_{\rm Ni} \lesssim 0.01\,M_\odot$.
Here we refer to two representative cases obtained by \cite{2022ApJ...927..223S} (see Table \ref{tbl:USSN}). The CO145 and CO20 models are for ultra-stripped progenitors with CO core masses of $1.45\,M_\odot$ and $2.0\,M_\odot$, respectively. The former (latter) represents a more (less) stripped progenitor and can be regarded to be in a binary system with a relatively small (large) orbital separation. The former (latter) shows a relatively strong (weak) explosion with a larger (smaller) ejecta mass. Noticeably, the latter has a significantly small $^{56}$Ni mass, which is mainly due to the fallback~\citep[see][for the details]{2022ApJ...927..223S}. 

\begin{table*}
\centering
    \caption{Properties of ultra-stripped supernova explosion given by \cite{2022ApJ...927..223S}.}\label{tbl:USSN}
    \begin{tabular}{ r c | c c c c | c c | c c c c c } \hline 
      Model & ${M_\mathrm{CO}}^{\rm a}$ & ${E_{\rm SN}}^{\rm b}$ & ${M_{\rm NS}}^{\rm c}$ & ${M_\mathrm{ej}}^{\rm d}$ & ${M_\mathrm{Ni}}^{\rm e}$ & ${\widetilde{{M}_\mathrm{fb}}}^{\rm f}$ & ${\widetilde{t_\mathrm{fb}}}^{\rm g}$ & ${X_\mathrm{C}}^{\rm h}$ & ${X_\mathrm{O}}^{\rm i}$ & ${X_\mathrm{Ne}}^{\rm j}$ & ${X_\mathrm{Si}}^{\rm k}$ & ${X_\mathrm{Fe}}^{\rm l}$\\
      & [$M_\odot$]& [$10^{51}\,\mathrm{erg}$] & [$M_\odot$] & [$M_\odot$] & [$M_\odot$] & [$10^{-2}\,M_\odot$] & [sec] & \\ \hline 
      CO145 & 1.45 & 0.17 & 1.35 & 0.097 & $1.63\times10^{-2}$ & $0.18$ & $10^3$ & $0.028$ & $0.33$ & $0.055$ & $0.22$ & $0.0057$ \\
      CO20 & 2.0& 0.12 & 1.64 & 0.286 & $7.78\times10^{-5}$ & $1.7$ & $10^3$ & $0.064$ & $0.51$ & $0.36$ & $0.073$ & $0.0020$\\
      \hline 
    \end{tabular}\label{tab:sim_res}
    
    \begin{tablenotes}[para,flushleft,online,normal] %(default:normal) 
    \item[a] Progenitor CO core mass;
    \item[b] Explosion energy;
    \item[c] NS mass;
    \item[d] Ejecta mass;
    \item[e] $^{56}$Ni mass;
    \item[f,g] Fitting parameters for the late-phase fallback (Eq. \ref{eq:fb_late});
    \item[h,i,j,k,l] Fractional abundance of carbon, oxygen, neon, silicon, and iron in the ejecta. 
    \end{tablenotes}

\end{table*}

On the basis of the results of the hydrodynamic simulations, the basic properties of USSN light curve can be inferred as follows.  
The optical depth of the ejecta evolves with time as $\tau_{\rm sc} \approx 3\kappa_{\rm sc} M_{\rm ej}/4\pi r_{\rm ej}^2$, or 
\begin{equation}
\tau_{\rm sc} \sim 13\,\kappa_{\rm sc, -0.7}M_{\rm ej,-1}v_{\rm ej, 9}^{-2}t_{5.9}^{-2},
\end{equation}
where $\kappa_{\rm sc} = 0.2\,{\rm cm^2\,g^{-1}}\,\kappa_{\rm sc, -0.7}$ is the electron scattering opacity and $v_{\rm ej} \sim 10^9\,{\rm cm\,sec^{-1}}\,E_\mathrm{sn,50}^{1/2} M_\mathrm{ej, -1}^{-1/2}$ is the velocity of the ejecta. 
The photon diffusion time through the ejecta is given by $t_{\rm dif} \approx \tau_{\rm sc} r_{\rm ej}/c$, or 
\begin{equation}\label{eq:t_dif}
t_{\rm dif} \sim 4.3\,{\rm day}\,\kappa_{\rm sc, -0.7}M_{\rm ej,-1}v_{\rm ej, 9}^{-1}t_{5.9}^{-1}.
\end{equation}
The SN light curve takes its peak when the diffusion time becomes comparable to the dynamical time of the ejecta, $t_{\rm dif}(t) \approx r_{\rm ej}(t)/v_{\rm ej}$. Solving this equation for time, the peak time can be estimated as $t_{\rm opt, peak} \approx (3\kappa_{\rm sc} M_{\rm ej}/4\pi c v_{\rm ej})^{1/2}$, or 
\begin{equation}\label{eq:tdif}
t_{\rm opt, peak} \sim 6.5\,{\rm day}\,\kappa_{\rm sc, -0.7}^{1/2}M_{\rm ej,-1}^{1/2}v_{\rm ej, 9}^{-1/2}.
\end{equation}
The energy injection rate by the ${}^{56}$Ni decay is 
\begin{equation}
\dot Q_{\rm Ni}(t) = M_{{\rm Ni}} \epsilon_{{\rm Ni}}\exp\left(-\frac{t}{t_{\rm Ni}}\right),
\end{equation}
where $\epsilon_{{\rm Ni}} = 3.9\times 10^{10}\,{\rm erg\,sec^{-1}\,g^{-1}}$ and $t_{{\rm Ni}} = 8.8\,\rm day$.
In the case of the USSN, $t_{\rm opt, peak} \lesssim t_{{\rm Ni}}$, and if the ${}^{56}$Ni decay is the main energy source, the peak luminosity can be estimated as  
\begin{equation}\label{eq:Loptp}
L_{\rm opt, peak} \approx \dot Q_{\rm Ni}(t_{\rm opt, peak}) \sim 7.8\times 10^{41}\,{\rm erg\,sec^{-1}}\,M_{{\rm Ni}, -2}.
\end{equation}

High-cadence photometric surveys are detecting rapidly evolving optical transients broadly consistent with Eqs. (\ref{eq:tdif}) and (\ref{eq:Loptp})~\citep[e.g.,][]{2014ApJ...794...23D,2016ApJ...819...35A,2016ApJ...819....5T}. Although a good fraction of them are likely explained by shock breakout or post-shock cooling emission from a dense circumstellar matter~\citep{2021arXiv210508811H}, some cases are spectroscopically confirmed to be compatible with ultra-stripped progenitors, e.g., iPTF14gqr~\citep{2018Sci...362..201D} and SN2019dge~\citep{2020ApJ...900...46Y}.  \cite{2022ApJ...927..223S} demonstrated that SN2019dge can be consistently explained by the neutrino-driven explosion of a more stripped progenitor as the CO145 model and the light curve powered by $^{56}$Ni decay. On the other hand, if iPTF14gqr is also powered by $^{56}$Ni decay, the required amount of ejected $^{56}$Ni mass is $M_\mathrm{Ni} \sim 0.05\,M_\odot$, which were shown to be difficult to synthesize by neutrino-driven explosion of ultra-stripped progenitors. This apparent $^{56}$Ni problem can be resolved by incorporating an additional energy injection into the USSN ejecta from the nascent NS~\citep{2016ApJ...819...35A,2022ApJ...927..223S} or CB.

\subsection{Nascent compact binary}
As a remnant of a USSN explosion in a close binary system, a BNS or a NS-BH is formed. Here, we consider the physical parameters of the nascent CB. 

If the orbital parameters satisfy Eq. (\ref{eq:a_crit}), the binary merges within a cosmological time and can be detected by the GW detector network. For GW events, the masses of compact objects are determined; $M_{\rm NS} \sim 1.5$ and $1.3\,M_\odot$ for GW170817~\citep{2017PhRvL.119p1101A}, $M_{\rm NS} \sim 2.0$ and $1.4\,M_\odot$ for GW190425~\citep{2020ApJ...892L...3A}, $M_{\rm BH} = 8.9^{+1.2}_{-1.5}\,M_\odot$ and $M_{\rm NS} = 1.9^{+0.3}_{-0.2}\,M_\odot$ for GW200105, and $M_{\rm BH} = 5.7^{+1.8}_{-2.1}\,M_\odot$ and $M_{\rm NS} = 1.5^{+0.7}_{-0.3}\,M_\odot$ for GW200115~\citep{2021ApJ...915L...5A}. Although the sample is limited, we assume that these masses are typical for NSs and BHs in compact binaries that coalesce within a cosmological timescale. In principle, spins of the compact objects can be simultaneously determined from the GWs, but so far the uncertainties are relatively large.

Other physical parameters of the NSs can be also inferred from observations of Galactic NSs. In Galactic BNSs with $t_{\rm GW} < 1\,\rm Gyr$, the first-born NSs have relatively weak magnetic fields ($B \sim 10^{9\mbox{-}10}\,\rm G$) and show fast rotation ($P_{\rm s} < 200\,\rm ms$)~\citep[][and references therein]{2016MNRAS.456.4089B};
%~\citep{2004Sci...303.1153L,2006Sci...314...97K,2010ApJ...722.1030W,2014ApJ...787...82F,2014MNRAS.443.2183F,2015ApJ...805..156S,2016ApJ...831..150L,2018ApJ...854L..22S,2018MNRAS.475L..57C}
the first-born NS has experienced a common envelope phase and a Case BB Roche robe overflow, it tends to evolve into a millisecond pulsar. %Due to this accretion process, the primary NSs may have slightly larger masses than the secondary NSs as observed in the Galactic BNSs~\citep[e.g.,][]{2004Sci...303.1153L}. 
The parameters of the second-born NSs at their birth are relatively uncertain; if they are similar to those of young NSs in the Galaxy, $B = 10^{12\mbox{-}14}\,\rm G$ and $P_{\rm s} = 10\mbox{-}100\,\rm ms$~\citep[e.g.,][]{2019RPPh...82j6901E}. %Unfortunately, Galactic NS-BHs are yet to be discovered. Probably in the SKA era? 
The weak natal kick of the second-born NS can still induce a non-negligible eccentricity of the binary orbit, as inferred by binary population synthesis calculations explaining the detected GW events~\citep[e.g.,][]{2022arXiv220106713K}. Observationally, the Galactic BNSs with $t_{\rm GW} < 1\,\rm Gyr$ have eccentricities ranging from $0.085 < e < 0.4$~\citep[][and references therein]{2016MNRAS.456.4089B}. Note that half of them have a relatively small value of $e < 0.2$, which is more compatible with the weak natal kick expected for USSNe.

\section{Supernova fallback onto nascent compact binary}\label{sec:fallback}
In an SN explosion, a fraction of the ejecta falls back onto the nascent compact object~\citep{1971ApJ...163..221C,1972SvA....16..209Z,1988Natur.333..644M}. 
In the case of neutrino-driven explosion, the fallback starts when the neutrino luminosity from the proto-NS significantly decreases, typically $\sim 10\mbox{-}100\,\mathrm{sec}$ after the onset of the core collapse, and is the most significant in the early phase; the total fallback mass is sensitive to the progenitor structure and the SN explosion dynamics~\citep[e.g.,][]{2016ApJ...818..124E,2016ApJ...821...69E,2016ApJ...821...38S,2022ApJ...926....9J}. In the later phase, the fallback rate becomes to follow the asymptotic relation; 
\begin{equation}\label{eq:Mdot}
\dot M_\mathrm{fb}(t) = \frac{2\widetilde{M_{\rm fb}}}{3\widetilde{t_{\rm fb}}} \left(\frac{t}{\widetilde{t_{\rm fb}}}\right)^{-5/3}. 
\end{equation}
Table \ref{tbl:USSN} includes the parameters characterizing the late-phase fallback ($\widetilde{M_{\rm fb}}$, $\widetilde{t_{\rm fb}}$) obtained for the CO145 and CO20 models by \cite{2022ApJ...927..223S}. The CO20 model shows a more intense fallback since it has a more dense core structure and a smaller explosion energy.
%Is there any study on the fallback in electron-capture SNe?

In the case of an USSN accompanying CB formation, the fallback accretion mode should change with time~(see Fig. \ref{fig:schem}). The important parameter is the fallback radius $r_{\rm fb} \approx (GM_{\rm NS}t^2)^{1/3}$ where the fallback material arriving at the central region at time $t$ starts falling toward the central region. For $r_\mathrm{fb} \lesssim a$, or $t\lesssim t_\mathrm{orb}$, where the fallback radius is smaller than the orbital separation of the nascent CB, the fallback is mainly on the second-born NS. %~\footnote{This stage can be further divided into $t_{\rm fb} \lesssim t \lesssim a/v_{\rm ej} \sim 400\,{\rm sec}\,a_{\rm 11.6}v_{\rm ej,9}$ and $a/v_{\rm ej} \lesssim t \lesssim t_{\rm orb}$; $r_{\rm fb}$, i.e., before and after the SN ejecta passes the primary compact object. The fallback is simply on the secondary NS in the former period while an accretion on the primary compact object also occurs and the fallback on the secondary NS is disturbed by the orbital motion in the latter period.}. 
The fallback material would not have sufficient angular momentum to form a disk around the second-born NS, thus directly accrete onto the NS magnetosphere or the surface~\citep[e.g.,][]{2021ApJ...917...71Z}. The mass accretion rate onto the NS should be comparable to the fallback rate, i.e., $\dot M_{\rm NS}(t) \approx \dot M_\mathrm{fb}(t) \propto t^{-5/3}$ for $t \lesssim t_\mathrm{orb}$. 
On the other hand, for $r_\mathrm{fb} \gtrsim a$, or $t \gtrsim t_{\rm orb}$, the fallback is on the binary system. The fallback material can be scattered by the binary orbit and on-average gain an angular momentum to form a circumbinary disk. The circumbinary disk mass further accretes with a viscous timescale~\citep[e.g.,][]{2014ApJ...783..134F},
\begin{equation}
t_\mathrm{vis} \approx \frac{1}{3\pi\alpha(h/r)^2} t_\mathrm{orb} \sim 10\,\mathrm{day}\,\left(\frac{\alpha}{0.1}\right)^{-1} \left(\frac{h/r}{0.1}\right)^{-2} a_{11}^{3/2} \left(\frac{m}{2.8\,M_\odot}\right)^{-1/2},
\end{equation}
where $\alpha$ is the viscous parameter and $h/r$ is the scale height of the disk. 
The time-averaged accretion rate from the circumbinary disk to the CB, $\dot M_\mathrm{CB}(t)$, behaves differently before and after $t \approx t_\mathrm{vis}$: For $t_{\rm orb} \lesssim t \lesssim t_\mathrm{vis}$, the accretion rate is regulated by the viscous timescale of the circumbinary disk, i.e., 
\begin{equation}\label{eq:fb_mid}
    \dot M_\mathrm{CB}(t) \approx \dot M_\mathrm{fb}(t) \frac{t}{t_\mathrm{vis}} = \frac{2\widetilde{M_{\rm fb}}}{3t_{\rm vis}} \left(\frac{t}{\widetilde{t_{\rm fb}}}\right)^{-2/3} \propto t^{-2/3} \ \ \ (t_\mathrm{orb} \lesssim t \lesssim t_\mathrm{vis}). 
\end{equation}
For $t \gtrsim t_\mathrm{vis}$, the accretion rate onto the CB is determined by the mass supply rate to the circumbinary disk, i.e., 
\begin{equation}\label{eq:fb_late}
    \dot M_\mathrm{CB}(t) \approx \dot M_\mathrm{fb}(t) \propto t^{-5/3} \ \ \ (t \gtrsim t_\mathrm{vis}).
\end{equation}
The accreted mass from the circumbinary disk forms mini disks around each compact object and finally accrete onto them. Similar situation, in particular accretion on a BBH, has been numerically investigated~\citep[e.g.,][]{2014ApJ...783..134F,2016MNRAS.459.2379D,2018PhRvL.120z1101T}.
For cases with a sufficiently large mass ratio $q = m_2/m_1 > 0.04$, the accretion rate on each compact object is on average comparable, but fluctuates with orbital motion~\citep{2014ApJ...783..134F,2016MNRAS.459.2379D}. 
%In our case, $q \sim 1$, thus we assume that the accretion rate on each NS is on average $\dot M/2$ and fluctuates with the orbital motion. 
We note that hardening of BNS / NS-BH by interacting with the fallback material is negligible since $M_{\rm fb, tot} \ll M_{\rm NS}$~\citep{2018PhRvL.120z1101T}.
We also note that the viscous timescales of the mini-disks are much smaller than those of both $t_\mathrm{orb}$ and $t_\mathrm{vis}$ and can be neglected for the estimate of $\dot M_\mathrm{CB}$. 

\begin{figure}
\centering
\includegraphics[width=0.49\textwidth]{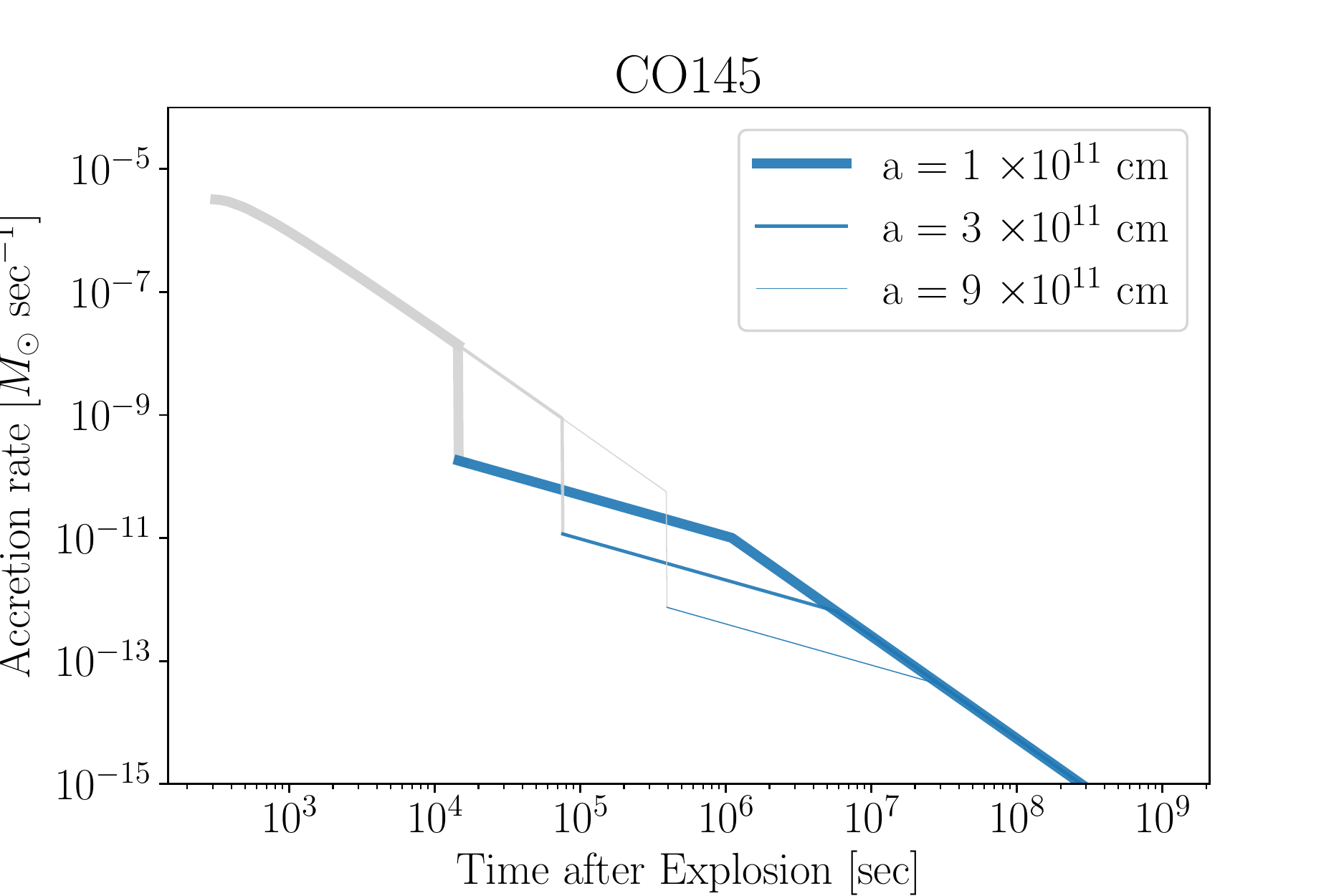}
\includegraphics[width=0.49\textwidth]{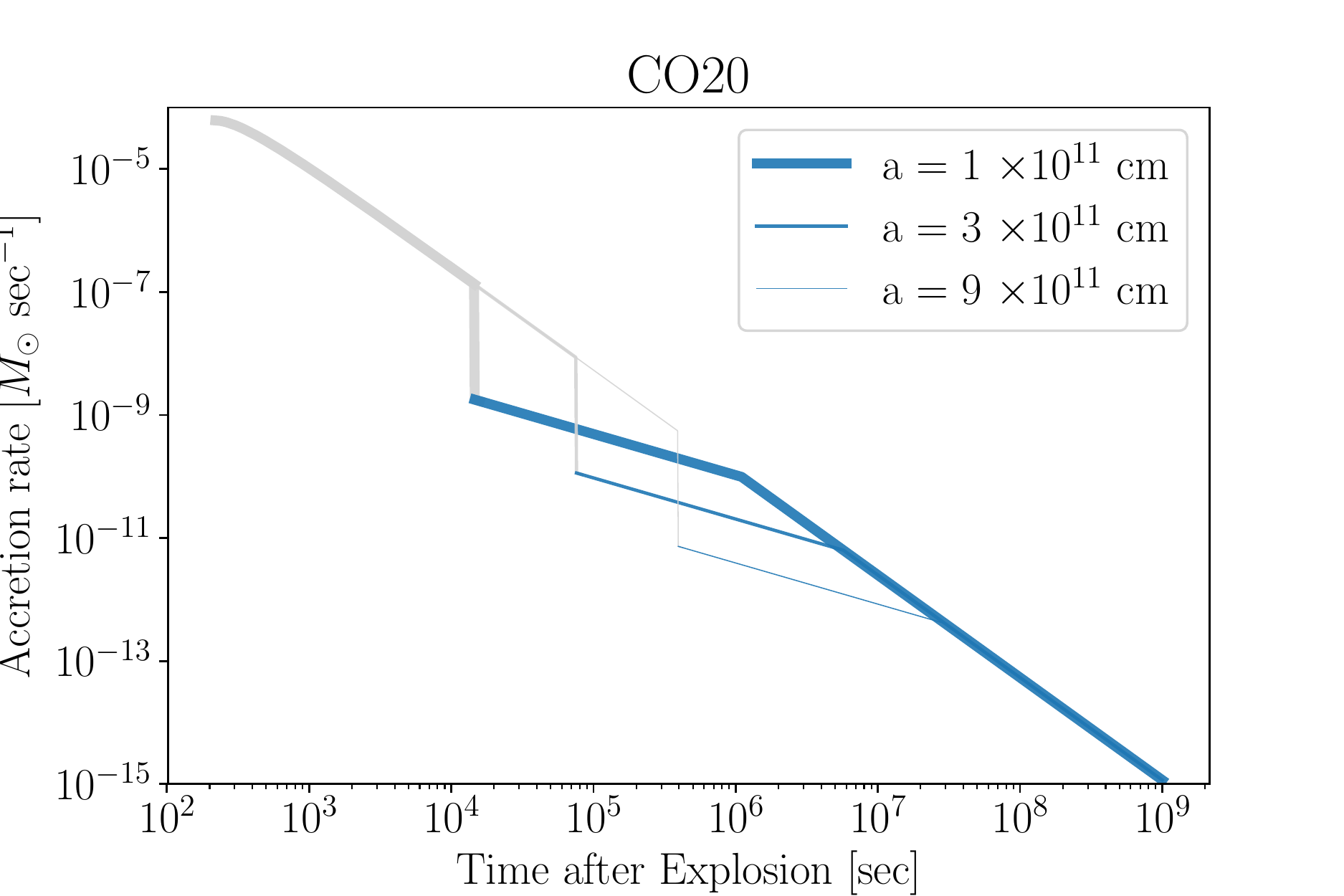}
\caption{
Fallback mass accretion rate onto a nascent compact binary system after a ultra-stripped supernova (USSN) at the birth of the second-born neutron star with an orbital separation of $a = (1, 3, 9) \times 10^{11}\,\mathrm{cm}$. The total mass of the binary is set to be $m = 2.8\,M_\odot$. The vertical lines indicate the orbital timescale and the viscous timescale of the circumbinary disk for the case with $a = 1\times 10^{11}\,\mathrm{cm}$. The USSN progenitor models are the CO145 (left) and CO20 (right) models in Table \ref{tab:sim_res}.
} 
\label{fig:acc}
\end{figure}

Figure \ref{fig:acc} shows the evolution of the time-averaged accretion rate onto the central compact objects based on the CO145 and CO20 models of \cite{2022ApJ...927..223S} with orbital separations of $a = 1\times 10^{11}\,\mathrm{cm}$, $3\times 10^{11}\,\mathrm{cm}$, and $9\times 10^{11}\,\mathrm{cm}$. We set $m = 2.8\,M_\odot$ , and assume $\alpha =0.1$ and $h/r = 0.1$ for the circumbinary disk.%, which would be reasonable given that the accretion rate is super-Eddington (see next section). 

\section{Binary ultraluminous X-ray source}~\label{sec:xray}
\begin{figure}
\centering
  \includegraphics[width=0.49\textwidth]{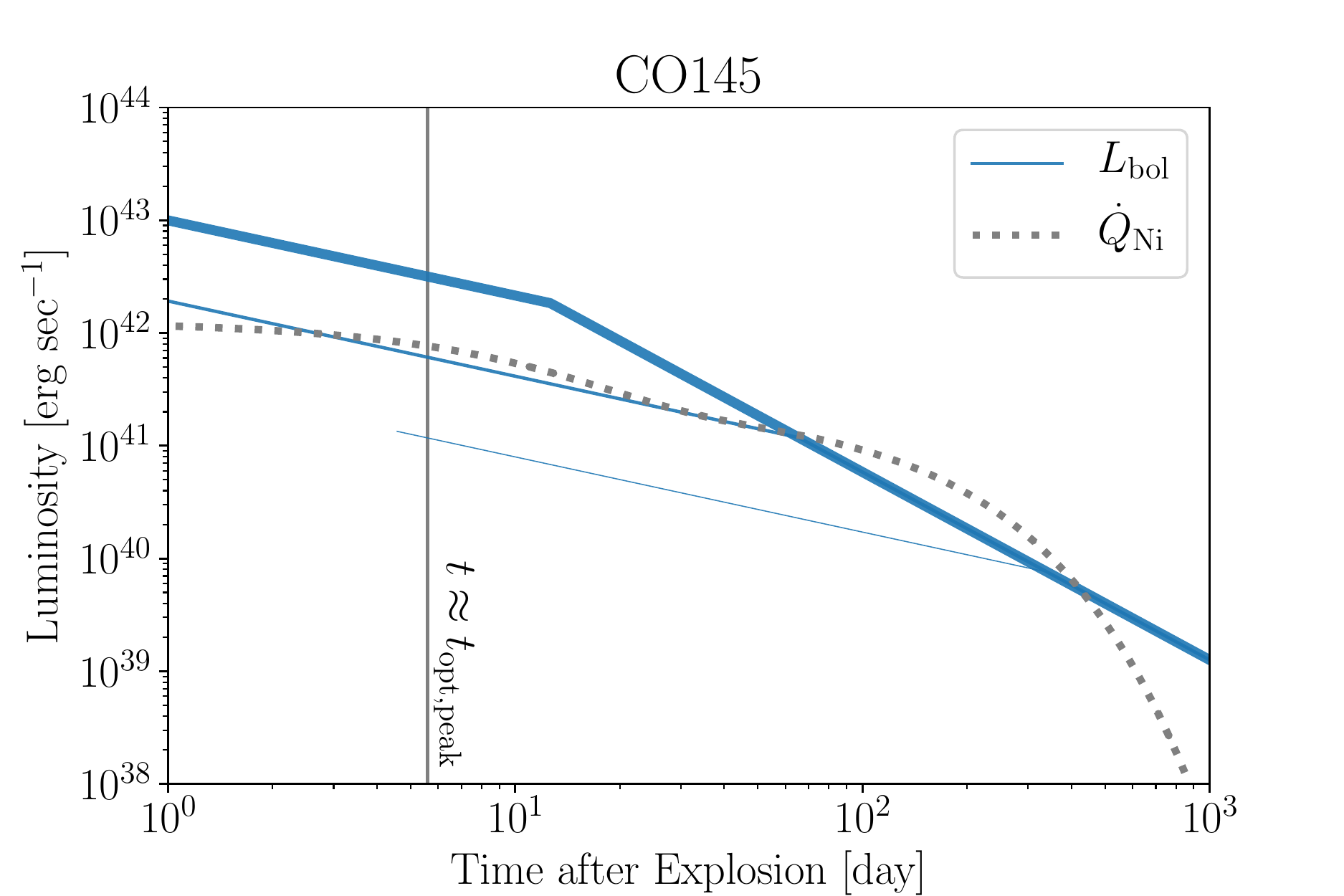} 
  \includegraphics[width=0.49\textwidth]{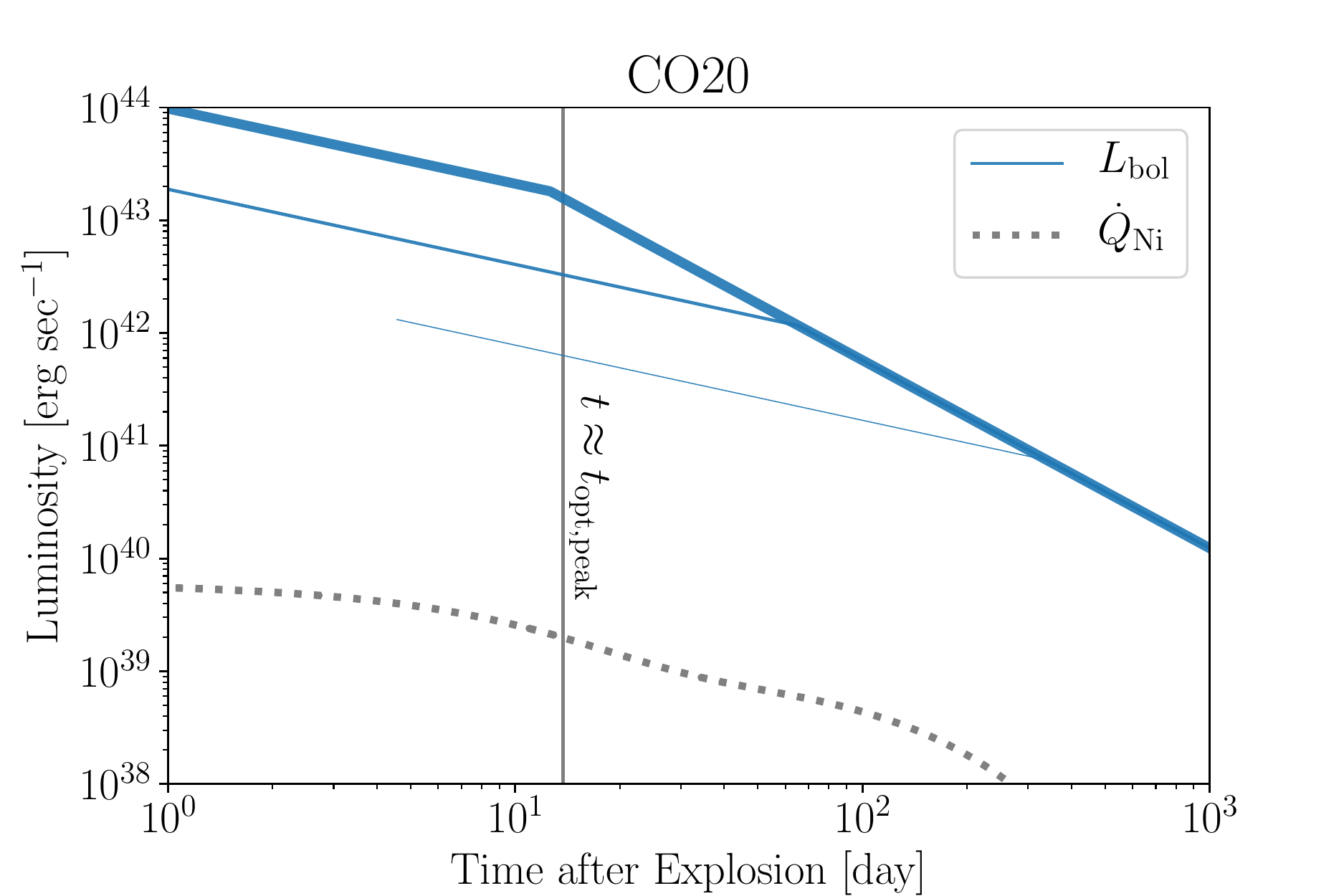} 
\caption{
Time-averaged bolometric luminosity of the accreting compact binary system with an radiation efficiency of $\eta = 0.1$ for the same models as Fig. \ref{fig:acc}. The dotted line show the energy injection rate to the USSN by the radioactive decay of $^{56}$Ni synthesised in the explosion. The vertical line indicates the calculated peak time of the optical light curve of the USSN. 
} 
\label{fig:L}
\end{figure}

Next let us consider emission from the accreting nascent CB. From Eqs. (\ref{eq:fb_mid}) and (\ref{eq:fb_late}), the average accretion rate normalized by the Eddington rate, $\dot M_{\rm Edd} \sim 4.0\times10^{-15}\,M_\odot\,{\rm sec^{-1}} \eta_{-1}^{-1} \kappa_{\rm sc, -0.7}^{-1} (m/2.8\,M_\odot)$, is described as 
\begin{equation}
\dot m \approx \frac{\dot M_\mathrm{CB}}{\dot M_{\rm Edd}} \sim 
\begin{cases}
1900 \,\eta_{-1}  \widetilde{M}_{\rm fb, -3} \widetilde{t}_{\rm fb, 3}^{2/3} \left(\frac{\alpha}{0.1}\right) \left(\frac{h/r}{0.1}\right)^2 a_{11}^{-3/2}\left(\frac{m}{2.8\,M_\odot}\right)^{-1/2} \kappa_{\rm sc, -0.7} t_{6}^{-2/3} \ \ \ (t_\mathrm{orb} \lesssim t \lesssim t_\mathrm{vis}), \\ 
36\,\eta_{-1} \widetilde{M}_{\rm fb, -3} \widetilde{t}_{\rm fb, 3}^{2/3} \left(\frac{m}{2.8\,M_\odot}\right)^{-1} \kappa_{\rm sc, -0.7}t_{7}^{-5/3} \ \ \ (t \gtrsim t_\mathrm{vis}). \\ 
\end{cases}
\end{equation}
where $\eta$ is the bolometric radiation efficiency. 
The accretion rate is higher than the Eddington rate up to
\begin{equation}
t_{\rm Edd} \sim 990\,{\rm day}\, \eta_{-1}^{3/5} \widetilde{M}_{\rm fb, -3}^{3/5} \widetilde{t}_{\rm fb, 3}^{2/5}\,\left(\frac{m}{2.8\,M_\odot}\right)^{-3/5} \kappa_{\rm sc, -0.7}^{3/5}.
\end{equation}
Fig. \ref{fig:L} shows the time evolution of the bolomotric luminosity of the accreting CB, $L_\mathrm{bol} = \eta \dot M_\mathrm{CO}c^2$, for the CO145 and CO20 models with $\eta = 0.1$. Note that $L_{\rm bol}$ includes contributions other than radiation, for example, the kinetic luminosity of the outflow. 
For $t_{\rm vis} \lesssim t \lesssim t_{\rm Edd}$, the accretion rate is comparable to the observed ULXs~\citep[see][for a recent raview]{2017ARA&A..55..303K}.
In this case, a dominant fraction of the accretion luminosity can be converted to X-rays, i.e. $\eta_{\rm X} \lesssim \eta$, and the luminosity can be described as 
\begin{equation}\label{eq:Lx}
\overline{L_{\rm X}} = \eta_{\rm X} \dot M_\mathrm{CB}c^2\sim 2.6 \times 10^{40}\,{\rm erg\,sec^{-1}}\,\eta_{\rm X, -1} \widetilde{M}_{\rm fb, -3} \widetilde{t}_{\rm fb, 3}^{2/3} t_{7}^{-5/3}, 
\end{equation}
which is the sum of the contributions from the two ULXs. The detail value of $\eta_{\rm X}$ and the spectral shape in the X-ray bands should depend on the physical properties of the accreting nascent CB, e.g., accretion rate, surface magnetic field strength and spin of the NS, and mass and spin of the BH.

As for the NSs, the observed properties of the ULX pulsars can be used as reference~\citep{2014Natur.514..202B,2016ApJ...831L..14F,2017Sci...355..817I,2017MNRAS.466L..48I}. The observed spectra of ULX pulsars are typically fitted by a double blackbody spectrum 
or a blackbody plus cutoff power law in an energy range of $\sim 1\mbox{-}10\,\rm keV$~\cite[e.g.,][]{2017A&A...608A..47K,2018ApJ...856..128W,2019ApJ...873...19T}. 
Although still under debate, the hard and soft components are considered to be coming from accretion column near the NS surface 
and accretion disk beyond the Alfv$\rm \grave{e}$n radius, respectively~\citep[see e.g.,][and references therein]{2019MNRAS.484..687M}.
The relative importance of each component is determined by the spin and magnetic field of the accretor.
The key parameters are the Alfv$\rm \grave{e}$n radius $R_{\rm M}$ where the magnetic pressure and the ram pressure of the accreting material balance, the co-rotation radius $R_{\rm co}$ where the rotation angular velocity of the NS and accretion disk becomes equal, and the spherization radius $R_{\rm sph}$ where the accretion luminosity reaches the Eddington limit~\citep{2018ApJ...856..128W}.   
They are estimated as 
\begin{equation}\label{eq:RM}
R_{\rm M} \approx \left(\frac{\mu_{\rm M}^4}{2 G M_{\rm NS}\dot M^2}\right)^{1/7} \sim 
\begin{cases}
1.4\times10^6\,{\rm cm}\,B_{10}^{4/7}\dot m_{2}^{-2/7} \left(\frac{M_{\rm NS}}{1.4\,M_\odot}\right)^{-1/7}, \\
7.3\times10^7\,{\rm cm}\,B_{13}^{4/7}\dot m_{2}^{-2/7}\left(\frac{M_{\rm NS}}{1.4\,M_\odot}\right)^{-1/7}, 
\end{cases}
\end{equation}
\begin{equation}\label{eq:Rco}
R_{\rm co} = \left(\frac{GM_{\rm NS}P_{\rm s}^2}{4\pi^2}\right)^{1/3} \sim 
\begin{cases}
1.6\times10^6\,{\rm cm}\,P_{\rm s,-3}^{2/3}\left(\frac{M_{\rm NS}}{1.4\,M_\odot}\right)^{1/3},\\
3.6\times10^7\,{\rm cm}\,P_{\rm s,-1}^{2/3}\left(\frac{M_{\rm NS}}{1.4\,M_\odot}\right)^{1/3}\\
\end{cases}
\end{equation}
\begin{equation}
R_{\rm sph, NS} \approx \frac{27}{4} \frac{\dot M}{\dot M_{\rm Edd}} \frac{GM_{\rm NS}}{c^2} \sim 1.4\times 10^{8}\,{\rm cm}\,\dot m_{2}\left(\frac{M_{\rm NS}}{1.4\,M_\odot}\right), \\
\end{equation}
respectively.
The upper and lower rows in Eqs. (\ref{eq:RM}) and (\ref{eq:Rco}) correspond to the first and second NS, respectively.

For the first-born NS, we expect $R_{\rm NS} \sim R_{\rm co} \sim R_{\rm M} \ll R_{\rm sph, NS}$; due to the relatively weak magnetic field, $R_{\rm M}$ becomes close to the surface and $\sim 100\,\dot m_2$ times smaller than $R_{\rm sph}$. In such case, the optically thick outflow should be relevant. Radiation efficiency may be suppressed to $\eta \gtrsim 1/\dot m$, that is, the total luminosity of X-rays becomes $\gtrsim L_{\rm Edd}$~\citep{1973A&A....24..337S}. However, even in this case, the apparent isotropic luminosity can be super-Eddington in a beamed direction~\citep[e.g.,][]{2016MNRAS.458L..10K,2017MNRAS.468L..59K}.
%Also, see \cite{2018ApJ...856..128W}.
For the second-born NS, we expect $R_{\rm co} \lesssim R_{\rm M} \lesssim R_{\rm sph, NS}$. In this case, the impacts of the optically thick outflow are relatively minor. The strong magnetic field may enable to maintain an accretion column near the surface and the radiation efficiency can be as high as $\eta_{\rm X} \sim 0.1$~\citep[e.g.,][]{2016PASJ...68...83K}. Since $R_{\rm co} \lesssim R_{\rm M}$, the second-born NS is likely spinning up. The parameters of the second-born NS are broadly consistent with those inferred for the observed ULX pulsars (although our fiducial spin period is slightly smaller than the observed values). 

For the first-born BH in a NS-BH, the accretion disk should be truncated at the innermost stable circular orbit (ISCO), $R_\mathrm{ISCO} \sim 4.5 \times 10^6\,\mathrm{cm}\,(M_{\rm BH}/5\,M_\odot)$. Other than the innermost structure, the basic properties can be similar to super-Eddington accretion to a first-born NS in a BNS; an outflow would be prominent within the spherization radius $R_{\rm sph, BH} \approx (27/4) \times (\dot M/\dot M_{\rm Edd}) \times (GM_{\rm BH}/c^2) \sim 5.0\times 10^8\,\mathrm{cm}\,\dot m_2 (M_{\rm BH}/5\,M_\odot)$, and the total X-ray luminosity may not be significantly larger than the Eddington limit, while the apparent isotropic X-ray luminosity can be super-Eddington in a beamed direction. The X-ray spectrum of the BH accretion disk in general consists of a multi-temperature disk component and a power-law component produced in the coronal region. In the cases with a high accretion rate we are interested in, the latter component would be minor, as observed in ULXs.

The X-ray emission from the accreting nascent CB intrinsically can have time variability. 
First, the emission may show modulation due to the orbital motion of the binary. 
Such a modulation has been detected from an accreting supermassive black hole binary candidate~\citep{2015Natur.518...74G}. 
Based on the numerical simulations~\citep[e.g.,][]{2014ApJ...783..134F}, the accretion rate on each compact object periodically fluctuates by more than a few 10\%,
which can be decomposed into evenly spaced frequencies, $\omega_i \approx 0.2 \times (2\pi/t_{\rm orb}) \times i$, where $i$ is positive integer. 
Second, the emission from near-surface regions, i.e., the accretion column and the inner disk, may show a (quasi)periodic variability associated with the secondary NS spin.
In the case of the observed ULX pulsars, the light curves are sinusoidal 
and the pulse fraction is $\sim$ 10\% in the soft X-ray band and slightly larger for the hard X-ray band~\citep{2014Natur.514..202B}. 
The spin period may decrease with time as the NS spins up due to the accretion. 
Third, the accretion dynamics of the mini disks can be imprinted in high-frequency quasi-periodic oscillations (QPOs) and low-frequency variability of the X-ray light curve~\citep[e.g.,][]{2006ARA&A..44...49R,2014Natur.513...74P}. In the BH case, the former would be correlated to the position of the ISCO and can be used to measure the BH mass.
%Also see \cite{2015Natur.525..351D}.

\section{X-raying nascent compact binaries?}\label{sec:xray_esc}

As we showed in the previous section, an accreting nascent compact binary in a USSN remnant can be a binary ULX, and the physical properties of the binary are imprinted in the X-ray emission. However, whether we can observe the signal is not straightforward. Just after the explosion, the SN ejecta is optically thick for electromagnetic waves.
For X-rays, inelastic Compton scattering and bound-free absorption are the main obstacles~\citep[e.g.,][]{2014MNRAS.437..703M}. 

The energy loss by inelastic Compton scattering is predominantly determined by the scattering optical depth; X-rays with an energy larger than 
\begin{equation}\label{eq:hsc}
(h\nu)_{\rm sc} \approx \frac{m_{\rm e}c^2}{\tau_{\rm ej}^2} \sim 3.1\,{\rm keV}\,\kappa_{\rm sc, -0.7}^{-2}M_{\rm ej,-1}^{-2}v_{\rm ej, 9}^{4}t_{5.9}^{4}
\end{equation}
will be lost in the ejecta. 
From Eq. \eqref{eq:hsc}, a good fraction of the X-ray luminosity is deposited to the USSN ejecta via inelastic scattering for $t \lesssim 10\,\rm day$, and can contribute to power the SN light curve. In Fig. \ref{fig:L}, we compare the bolometric luminosity with the decay rate of $^{56}$Ni synthesized and ejected in the explosion. In the case of a more stripped progenitor (CO145), the deposition of the X-rays and / or the kinetic luminosity of the outflow can give a comparable contribution to the $^{56}$Ni decay, while in the case of a less stripped progenitor (CO20), the deposition can be the main energy source of the USSN. In particular, the peak luminosity of the USSN can be even much higher than that of the canonical core-collapse SNe when the orbital separation is relatively small. This case may be applicable to the bright end of rapidly evolving optical transients~\citep[e.g.,][]{2020ApJ...895L..23C}. Detailed calculations of the optical light curve powered by the energy injection from the nascent CB and comparison with the observed transients will be presented elsewhere (Sawada et al., in prep). 

Hereafter, we focus on the transmitted X-ray emission through the USSN ejecta at $t \gtrsim 10\,\mathrm{day}$. We pick up two cases, the CO145 model with $a = 1\times 10^{11}\,\mathrm{cm}$ and the CO20 model with $a = 9\times 10^{11}\,\mathrm{cm}$, for which the optical light curve will be broadly consistent with a typical USSN. 
For $t \gtrsim 10\,\mathrm{day}$, bound-free absorption~\citep[e.g.,][]{2006agna.book.....O} may still be important. The threshold energy for the absorption is given by
\begin{equation}\label{eq:bf}
(h\nu)_{\rm bf} \approx 870\,{\rm eV}\,Z_8^2.
\end{equation}
where $Z = 8\,Z_8$ is the average degree of ionization of the ejecta. The ionization rate is given by $\Gamma_{\rm ion} \approx n_\gamma \sigma_\gamma c$, or
\begin{equation}\label{eq:tion}
\Gamma_{\rm ion} \sim 0.76\,{\rm sec^{-1}}\,Z_8^{-4} \overline{L_{\rm x}}_{,41} v_{\rm ej,9}^{-2} t_{5.9}^{-2},
\end{equation}
where $n_\gamma = \overline{L_{\rm x}}/[4\pi c r_{\rm ej}^2 (h\nu)_{\rm bf}]$ is the ionizing photon density and $\sigma_\gamma \sim 1\times 10^{-19}\,{\rm cm^2}\,Z_8^{-2}$ is the cross section. On the other hand, the recombination rate is given by $\Gamma_{\rm rec} \approx n_{\rm e}\alpha_{\rm rec}$, or 
\begin{equation}\label{eq:trec}
\Gamma_{\rm rec} \sim 0.44\,{\rm sec^{-1}}\,Z_8^2T_4^{-0.8} M_{\rm ej,-1}^{-1} v_{\rm ej,9}^{-3} t_{5.9}^{-3},
\end{equation}
where $n_{\rm e} = 3 M_{\rm ej}/8\pi m_{\rm u} r_{\rm ej}^3$ is the electron density and $\alpha_{\rm rec} \sim 2\times 10^{-11}\,{\rm cm^3\,sec^{-1}}\,Z_8^2 T_{\rm e,4}^{-0.8}$ is the (case B) recombination coefficient. The electron temperature should be determined including the Compton heating effect, ranging from $T_{\rm e} \sim 10^{4\mbox{-}5}\,\rm K$. From Eqs. \eqref{eq:Lx}, \eqref{eq:tion}, and \eqref{eq:trec}, the condition $\Gamma_{\rm ion} > \Gamma_{\rm rec}$ is realized for 
\begin{equation}\label{eq:Zcri}
Z < 11\,T_{\rm e,4}^{2/15} M_{\rm ej,-1}^{1/6} v_{\rm ej,9}^{1/6}\eta_{-1}^{1/6} \widetilde{M}_{\rm fb, -3}^{1/6} \widetilde{t}_{\rm fb, 3}^{1/9} t_{\rm 5.9}^{-1/9}.
\end{equation}
Eq. (\ref{eq:Zcri}) may infer that elements up to neon can be fully ionized by the X rays. To calculate the transmitted X-ray spectrum through the USSN ejecta, we need to solve the ionization state of heavier elements and the transfer of the radiation field consistently. 

Here we use the spectral synthesis code \texttt{CLOUDY}~\citep{1998PASP..110..761F,2017RMxAA..53..385F}. For simplicity, we fix the shape of the injection X-ray spectrum as a multi-temperature blackbody with a temperature range of $T_\mathrm{bb} \sim (1\mbox{-}3) \times 10^7\,\mathrm{K}$, mimicking the observed ULX spectra. The total flux is determined from Eq. (\ref{eq:Lx}). We assume a homologously expanding ejecta; the density profile and composition are set to be consistent with those obtained by hydrodynamic simulations~\citep{2022ApJ...927..223S}. The fractional abundances of some relevant elements are shown in Table \ref{tbl:USSN}. We neglect the interstellar absorption both in the host galaxy and in the Milky way. 

\begin{figure}
\centering
  \includegraphics[width=0.49\textwidth]{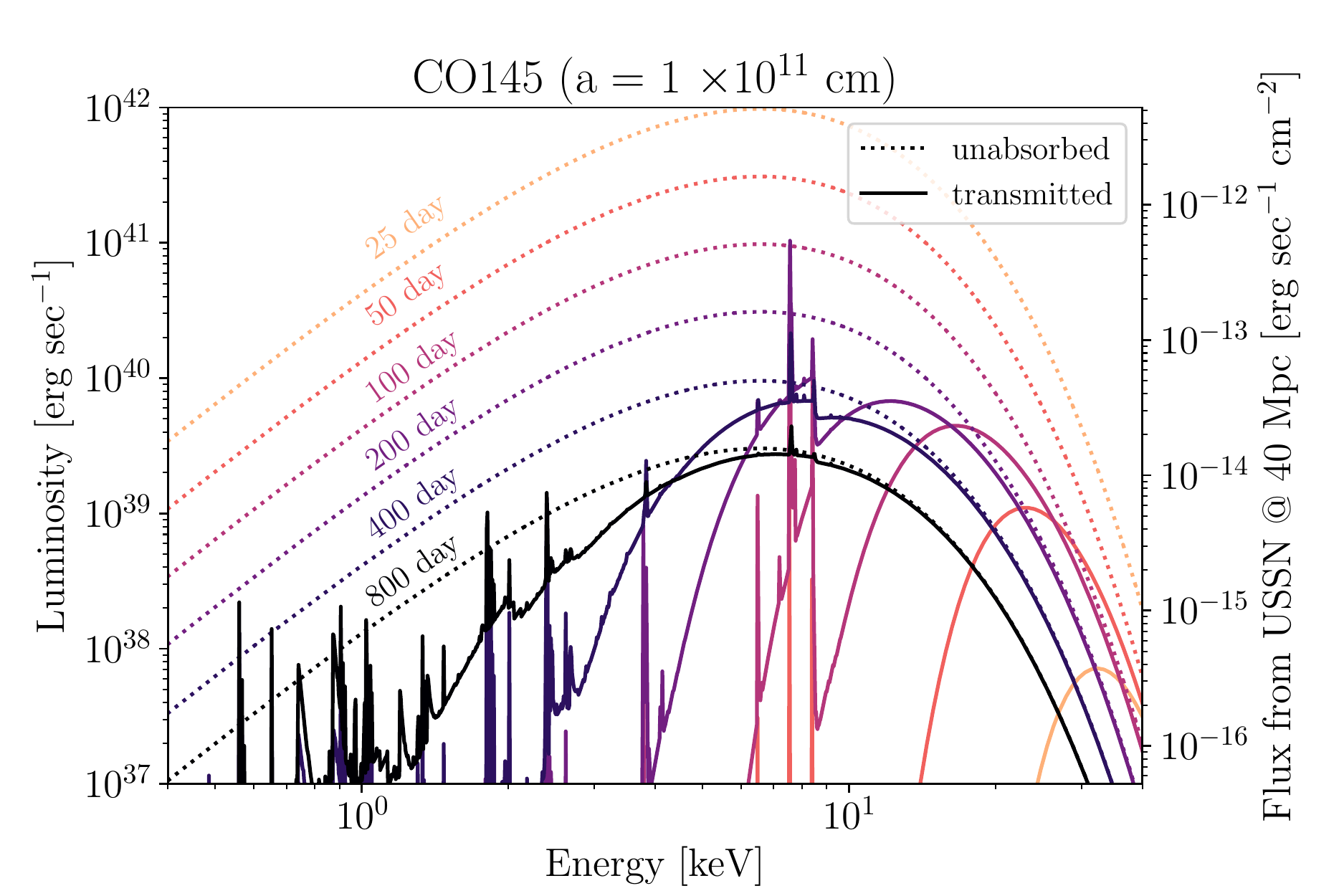}
  \includegraphics[width=0.49\textwidth]{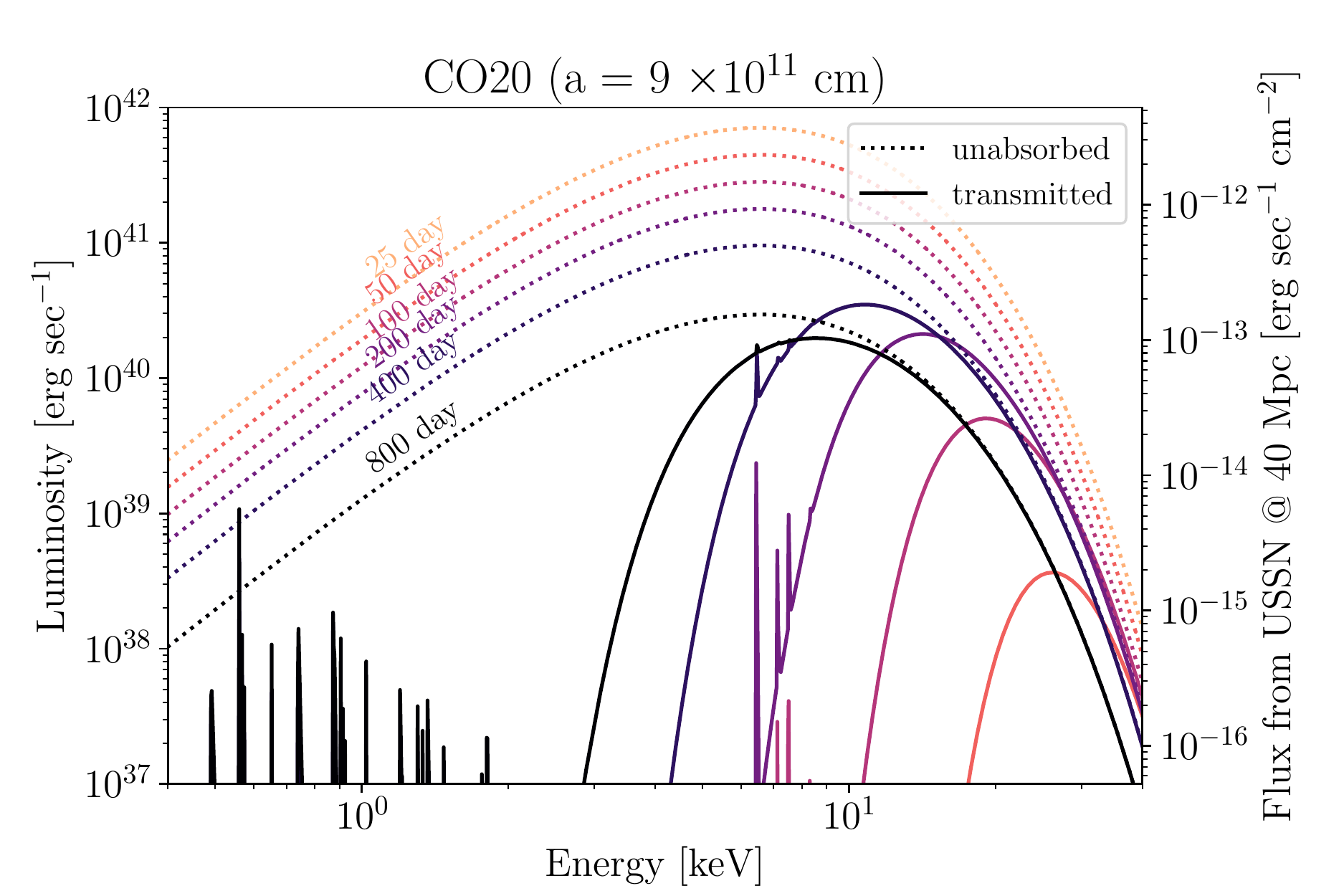}
\caption{
Time evolution of X-ray spectrum of nascent compact binaries formed with explosion of ultra-stripped progenitors, the CO145 model with an orbital separation of $a = 1\times 10^{11}\,\mathrm{cm}$ (left) and the CO20 model with $a = 9\times 10^{11}\,\mathrm{cm}$ (right). 
The dotted lines show the injection spectra assuming to be a multi-temperature black body and the solid lines show the transmitted spectra through the supernova ejecta at 25, 50, 100, 200, 400, 800 day after the explosion. 
} 
\label{fig:spec}
\end{figure}

Fig. \ref{fig:spec} shows the resultant X-ray spectra of the CO145 model with $a = 1\times 10^{11}\,\mathrm{cm}$ and the CO20 model with $a = 9\times 10^{11}\,\mathrm{cm}$ at 25, 50, 100, 200, 400, 800 days after the explosion. In both cases, the emitted X-rays from the accreting nascent CB are absorbed significantly at an early stage, but gradually escape through the ejecta from $\sim$ one month after the explosion. The ejecta becomes transparent first for hard X-rays and then for intermediate to soft X-rays. We note that the recombination lines of the iron group elements are less prominent in the CO20 model, since the abundance is lower (see Table \ref{tbl:USSN})~\footnote{Note that we neglect the ejecta expansion in the calculation of the transmitted spectra, thus the actual emission lines will be broader with a typical width of $\Delta E/E \approx v_\mathrm{ej}/c \sim 0.03\,v_\mathrm{ej,9}$, which could be probed by high-resolution X-ray spectroscopy with XRISM~\citep{2020SPIE11444E..22T} and Athena~\citep{2018SPIE10699E..1GB}.}.

\begin{figure}
\centering
  \includegraphics[width=0.49\textwidth]{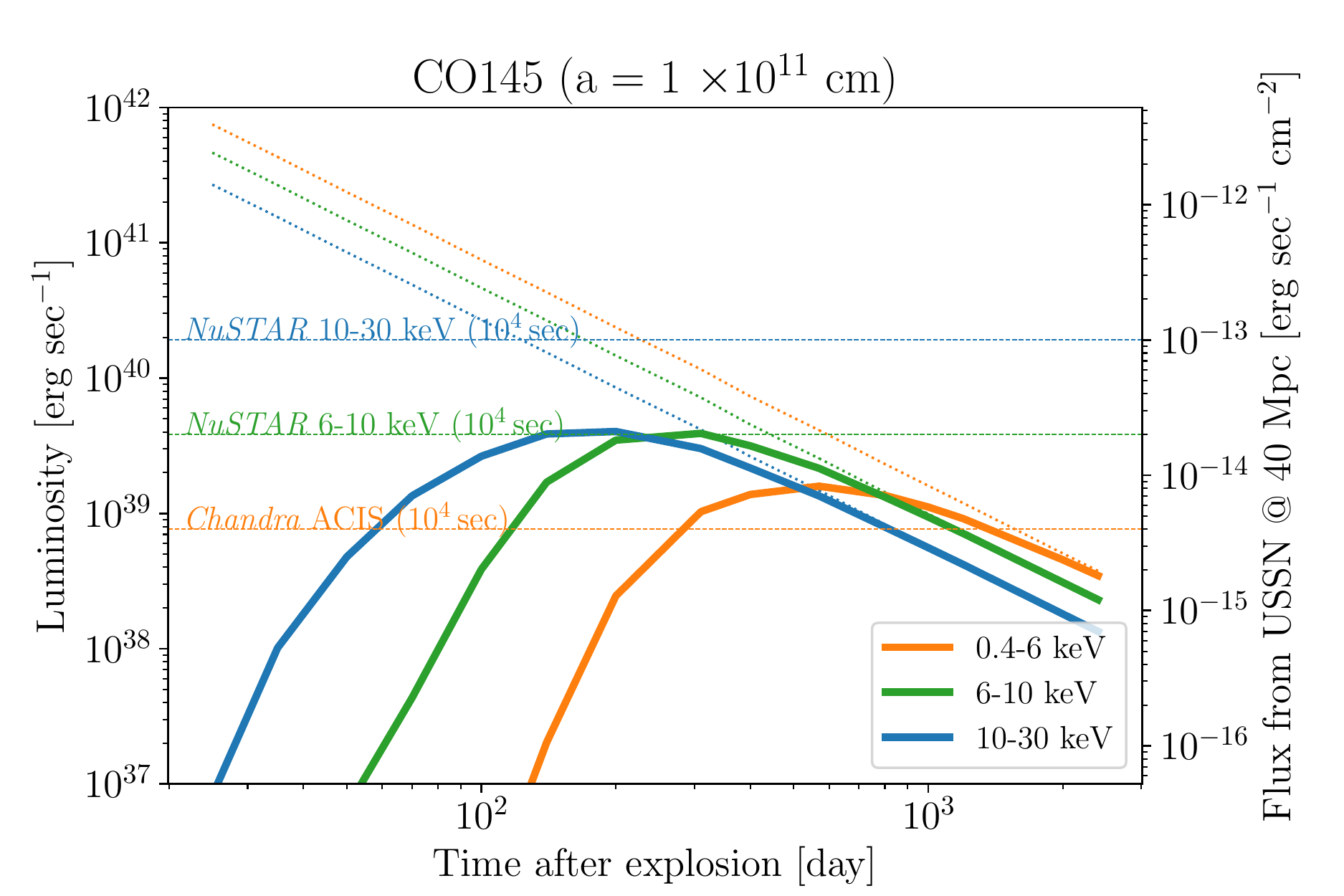}
  \includegraphics[width=0.49\textwidth]{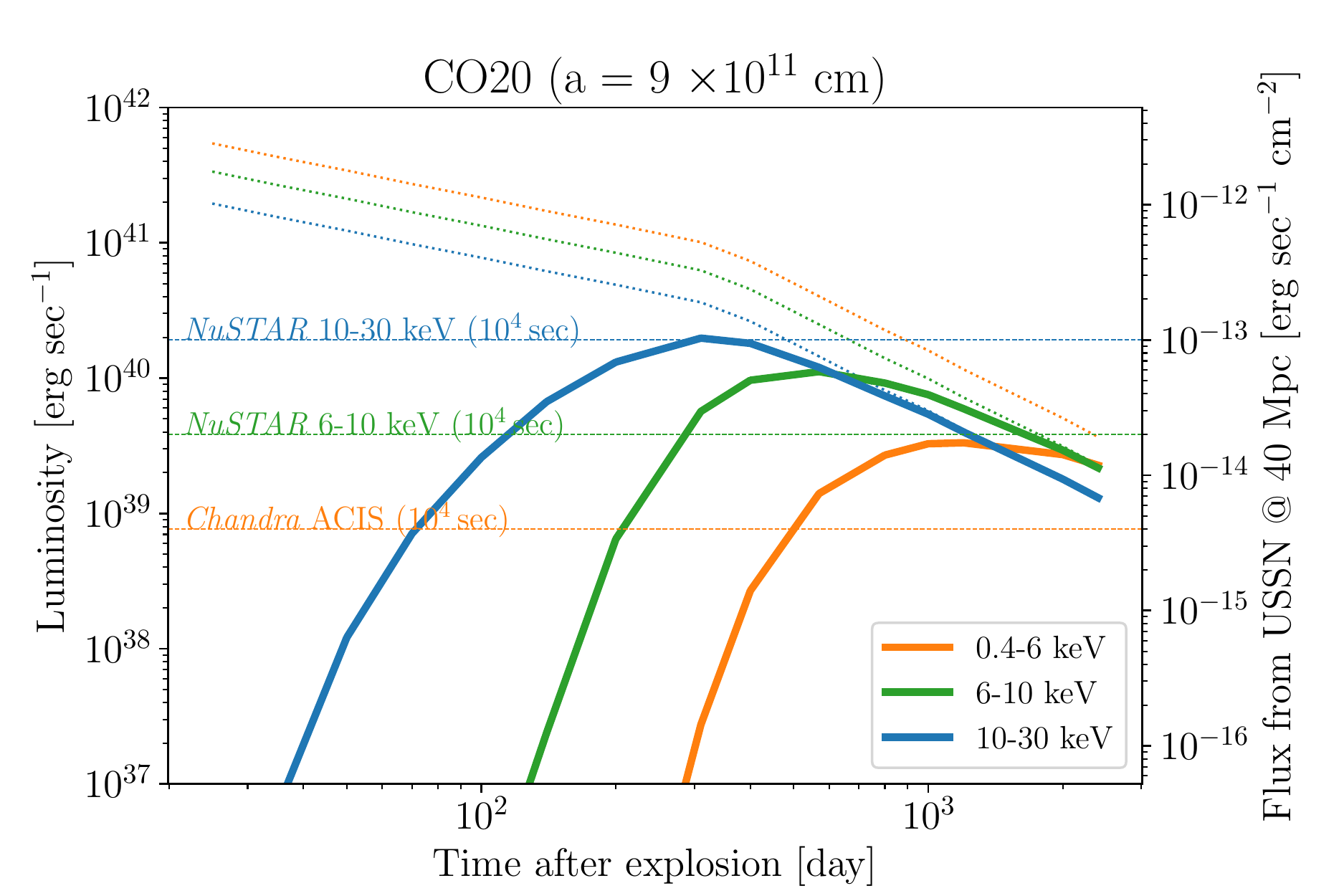}
\caption{
X-ray light curves of the same nascent compact binary models as Fig. \ref{fig:spec}. The dotted lines show the injection fluxes from the accreting compact binaries and the solid lines show the transmitted fluxes through the supernova ejecta for the soft ($0.4\mbox{-}6\,\mathrm{keV}$), intermediate ($6\mbox{-}10\,\mathrm{keV}$), and hard ($10\mbox{-}30\,\mathrm{keV}$) X-ray bands.  
} 
\label{fig:lc}
\end{figure}

Fig. \ref{fig:lc} shows the corresponding light curves in the soft ($0.4\mbox{-}6\,\mathrm{keV}$), intermediate ($6\mbox{-}10\,\mathrm{keV}$), and hard X-ray ($10\mbox{-}30\,\mathrm{keV}$) bands. For comparison, we indicate the sensitivities of {\it Chandra} ACIS and {\it NuSTAR} with an integration time of $10^4\,\mathrm{sec}$ for a USSN at a nominal distance of $40\,\mathrm{Mpc}$. The X-ray counterpart in the soft and intermediate bands can be detectable up to $\lesssim 100\,\mathrm{Mpc}$ by these instruments (and {\it XMM Newton} with a sensitivity comparable to {\it Chandra}). The detection will also be promising for future X-ray satellites, e.g., {\it XRISM}~\citep{2022arXiv220205399X} and {\it Athena}~\citep{2013arXiv1306.2307N}. The expected event rate of such USSNe is a few $\mathrm{yr^{-1}} \times f_\mathrm{b}$ where $f_\mathrm{b} \sim 0.1\mbox{-}1$ is the beaming fraction of the X-ray emission. The hard X-ray counterpart can also be detected for cases with a larger fallback rate and/or a smaller distance to the source. To this end, follow-up observations of USSNe need to be done within $\lesssim$ a few $100\,\mathrm{day}$ after the explosion in the intermediate and hard X-ray bands and within $\lesssim 1,000\,\mathrm{day}$ in the soft-X-ray band.
%A good spatial resolution may be also required.

Time variability of the accreting nascent CB can be also detected if the signal-to-noise ratio (S/N) is sufficiently high: In order to detect an abrupt variation with an amplitude of 10(1)\% and a duration of $\Delta t$, an S/N $\gtrsim$ 10(100) needs to be obtained with an integration time of $\sim \Delta t$, while if the variation is periodic, the S/N for the detection can be reduced by a factor of $\sim (\Delta t/T_\mathrm{obs})^{1/2}$, where $T_\mathrm{obs}$ is the total observation time. The most promising target would be the orbital modulation with a period of $t_\mathrm{orb} \sim 0.1\mbox{-}1\,\mathrm{day}$. The rotation period and its time derivatives of the second-born NS and the QPOs of the mini disks could also be detected when the signals are relatively persistent in time. 
%The diffusion timescale through the ejecta needs to be smaller than the variability timescale. In our case, $\tau_{\rm sc} < 1$ otherwise $t_{\rm dif} \lesssim P_{\rm spin}$ or $t_{\rm orb}$. 

\section{Summary and Discussion}\label{sec:summary}
We have investigated the effect of USSN fallback occurring after the formation of the secondary NS in a BNS or a NS-BH that coalesces within a cosmological timescale. We showed that the nascent CB can be a binary ULX and the X-ray counterpart is detectable by a followup observation of USSNe within $\lesssim 100\,\mathrm{Mpc}$ and $\sim 100\mbox{-}1,000\,\mathrm{day}$ after the explosion using {\it Chandra}, {\it XMM Newton}, {\it NuSTAR}, and future X-ray satellites, which provides a direct evidence of CB formation in USSN. Furthermore, information on the nascent CB, e.g. the orbital separation, eccentricity, and the NS spin, can be obtained from the time variability of the X-ray light curve. 
%In fact, such an X-ray follow-up observation has been conducted for iPTF16asu at $z = 0.187$~\citep{2017ApJ...851..107W}. 
The conclusions are based on simplified calculations of the fallback accretion on and the X-ray emission from the nascent CB and need to be justified by multidimensional radiation hydrodynamic simulations.

For $t \lesssim 100\,\mathrm{day}$, the X-ray counterpart could not be observed; though the accretion rate is even higher, the X-ray radiation efficiency may not be high and the X-rays, if any, are absorbed and converted into UV and optical photons. Instead, the injection of energy from the accreting nascent CB can power the USSN light curve, which may solve the apparent $^{56}$Ni problem raised for, e.g., iPTF14gqr~\citep{2022ApJ...927..223S}. The detailed modeling and comparison with observed USSNe and rapidly evolving optical transients will be presented elsewhere. 
%{\bf The observed bump in the optical light curve of SN2019dge may be explained by the onset of the bound-free heating. The timing is consistent.}

We finally note that, if the second-born NS is strongly mangetized and rapidly rotating, the spindwon luminosity can be comparable to or larger than the total accretion luminosity of the nascent CB. It can both provide an enhanced USSN light curve~\citep{2017ApJ...850...18H,2022ApJ...927..223S} and a late phase X-ray emission~\citep{2014MNRAS.437..703M,2016ApJ...818...94K}. To distinguish the energy sources, multi-wavelength follow-up observations also in the radio, submillimeter, and gamma-ray bands would be useful. 
%A similar followup strategy may work for searching nascent BBHs.

\section*{Acknowledgments}
KK thanks Aya Bamba, Daichi Tsuna, Toshikazu Shigeyama, and Shigeo S. Kimura for discussion. 
This work is financially supported by JSPS KAKENHI grant 18H04573 and 17K14248 (K.K.), 18H05437, 20H00174, 20H01904, and 22H04571 (Y.S.), 21J00825 and 21K13964 (R.S.).

\bibliography{ref}{}
\bibliographystyle{aasjournal}

\end{document}